\documentclass[10pt,letterpaper,twocolumn]{article}
\usepackage[utf8]{inputenc}
\usepackage{cite}
\usepackage{nameref,hyperref}
\usepackage{amsmath}
\usepackage{microtype}
\DisableLigatures[f]{encoding = *, family = * }

\usepackage{graphicx}

\begin{document}
\onecolumn

\title{The Excess Proton at the Air-Water Interface: The Role of Instantaneous Liquid Interfaces.}

\maketitle
Federico Giberti(1),
Ali A. Hassanali(2),\\

(1) The University of Chicago, Institute For Molecular Engineering, 5640 South Ellis Avenue, 
Chicago, Illinois 60637, USA.

(2) The Abdus Salam International Centre for Theoretical Physics, Condensed Matter and Statistical 
Physics, Strada Costiera 11, 34151 Trieste, Italy
* gibertif@uchicago.edu

\section*{Abstract}
The magnitude of the pH of the surface of water continues to be a contentious topic in the physical chemistry of aqueous interfaces. Recent theoretical studies have shown little or no preference for the proton to be at the surface compared to the bulk\cite{baer2014toward}. Using ab-initio molecular dynamics simulations, we revisit the propensity of the excess proton for the air-water interface with a particular focus on the role of instantaneous liquid interfaces. We find a much a stronger propensity of the proton for the surface of water. The enhanced water structuring around the proton results in the presence of proton wires that run parallel to the surface as well as a hydrophobic environment made up of under-coordinated topological defect water molecules, both of which create favorable conditions for proton confinement at the surface. The Grotthuss mechanism within the structured water layer involves a mixture of both concerted and closely spaced stepwise proton hops. The proton makes excursions within the first solvation layer either in proximity to or along the instantaneous interface. 

\twocolumn
\section*{Introduction}
Water interfaces are ubiquitous in nature - the cytoplasmic
membrane\cite{bellissent2016water}, solid catalysts\cite{chemrevinterfaces2016} and artificial photosynthetic devices\cite{chemrevphotosynthesis2014} are a few examples of practical
importance where these interfaces occur. The properties of these systems are heavily dictated by the physical and chemical interactions of water under these conditions. One of the most studied systems in this regard is the air-water
interface\cite{ChemRev2,ChemRev3} since it serves as a model system for
studying water near hydrophobic surfaces. There is an increasing effort from both experimental and theoretical fronts to understand the structural and
dynamical properties of the air-water interface since the literature
in the area is replete with contradictions, specifically regarding the value of the pH at the interface\cite{ChemRev3,buch2007water,Beattie2008,BeattieDjerdjevWarr2009}.

There are two main competing views regarding the pH of the surface of water. Numerous electrokinetic experiments\cite{Beattie2008,BeattieDjerdjevWarr2009} over the last couple of decades, and more recently electrospray ionization experiments probing chemical reactivity at the surface of water\cite{Mishra2012}, conclude that this interface is basic. According to this interpretation, hydroxide ions would have a high propensity to bind to the surface of water. On the other hand, sum frequency generation (SFG)\cite{TarbuckOtaRichmond2006} and second harmonic generation (SHG)\cite{petersensaykally2005,PetersenSaykally2008} have pointed to a surface that is proton enriched. More recently, another view has been put forward suggesting that the surface of water is neutral and that the negative zeta potential measured in electrokinetic experiments stems from a charge-transfer mechanism between water molecules\cite{soniat2015hydrated}.

Theoretically, the affinity of protons and hydroxide ions for the interface has typically been studied using molecular dynamics simulations based on Density Functional Theory (DFT)\cite{baer2014toward,buch2007water,MundyKuoTuckermanLeeTobias2009,jungwirthproton2011}, the Multi-State Empirical Valence-Bond method (MS-EVB)\cite{WuChenWangPaesaniVoth2008,KnightVoth2012,TseLindbergKumarVoth2015,wick2012hydronium} and finally force-field base approaches\cite{chenherzfield2016}. In these studies, the ions affinity for the surface is determined by constructing a one-dimensional potential of mean force (1d-PMF) as a function of the distance of the hydronium ion from the Gibbs dividing interface (GDI). Recent DFT simulations suggested little or no preference for the proton to bind to the surface\cite{baer2014toward}. Earlier MS-EVB studies showed a stronger propensity for the hydronium to be stabilized at the air-water interface compared to more recent investigations\cite{WuChenWangPaesaniVoth2008}.

Water in contact with the vapor phase is a lot more structured than that inferred by an analysis built on the GDI. This feature was demonstrated by Willard and Chandler who introduced the notion of instantaneous liquid interfaces (WCI)\cite{WillardChandler2010,kuhne2015}. In this work, we re-visit the propensity of the proton for the air-water interface with a particular focus the role of the WCI on both the thermodynamic and dynamical properties of the proton at the surface of water using DFT-based ab-initio molecular dynamics (AIMD) simulations. We find evidence for a much stronger propensity of the proton for the surface than that inferred from the one-dimensional PMFs built exclusively on the Euclidean distance of the proton from the GDI consistent with a very recent study examining the PMF with respect to the WCI\cite{Wick2017}. Our simulations elucidate that the origin of this feature is rooted in the structured water around the proton leading to very particular topological traits of its surrounding hydrogen bond network. In particular, we show that preferred orientation of water molecules leads to proton wires running parallel to the surface which biases the direction of proton motion which we suggest is rooted in the slow residence time exchange of water between the bulk and air-water interface. The proton trapping bears interesting similarities to what is observed near hydrophobic surfaces like at the surface of membrane proteins\cite{Pohl2012} as well as near metal surfaces like platinum\cite{Limmer2013}. 

We also elucidate mechanistic aspects of Grotthuss structural diffusion of the proton at the interface and find that it involves the collective reorganization of structured water molecules within the water layer identified by the WCI analysis. Despite being confined at the interface, proton motion undergoes excursions where it can \emph{surf} close to the WCI 
and makes rather large fluctuations within the first solvation layer of the interface. The proton hopping events involve a mixture of both stepwise and concerted hopping events which is coupled to subtle changes in the local hydrophobic environment around the proton similar to what is observed for proton transfer in the bulk\cite{HassanaliGibertiCunyKuhneParrinello2013,HassanaliGibertiSossoParrinello2014,KnightVoth2012}.

\section*{Materials and Methods}
\label{sec:method}

Extensive AIMD simulations of an excess proton at the air-water interface were conducted. The initial configurations into which the excess proton was inserted at the surface and sub-surface were obtained from previously published simulations of the neat air-water interface by K\"uhne and co-workers\cite{kuhne2011} equilibrated for over 200 ps. The system consists of a slab of 384 water molecules where the x, y and z dimensions are 15.6404, 85.0000, 15.6404\AA{} respectively. The air-water interface lies in the y-direction where approximately 40.00 \AA{} of vacuum buffer separated the two surfaces. 

We performed all the simulations with the CP2K suite using Quickstep\cite{VandeVondeleKrackMohamedParrinelloChassaingHutter2005}. The wavefunction was expanded using both a TZV2P Gaussian basis set as well as a plane wave representation with a cutoff of 320 Ryd.  The BLYP exchange-correlation was used together with the Grimme D3 empirical corrections for the van-der-Waals interactions\cite{grimme2010consistent}. We integrated the MD simulations with a 
timestep of 0.5 fs within the NVT ensemble using the CSVR thermostat\cite{bussi2007canonical}, thermostating the system at 300 K. 

We collected a total of over 130 ps with the proton lying at the water interface as well as below it. As alluded to earlier, previous studies examined the propensity of the proton for the surface calculating the PMF along the direction perpendicular to the air-water interface. In these simulations, the proton is locally constrained and is not allowed to undergo Grotthuss diffusion. In our studies, we do not compute the PMF but instead, evolve the proton from various initial conditions enabling us to probe the mechanistic details of the Grotthuss mechanism at the surface of water.  We initiated two trajectories where the proton was at the surface of water (labeled A and B) and another two with the proton at 15 \AA{} from the center of the slab (labeled C and D). In all simulations (A-D), the system was equilibrated by constraining the coordination number of a particular tagged oxygen to a value of 3 using a harmonic potential with a force constant of 0.1 a.u. These constrained simulations were run for a total of 20 ps before launching the unconstrained trajectories. The various simulations performed along with the initially constrained position of the excess proton perpendicular to the surface is shown in Table \ref{tab:tab1}.

To calculate the residence time of the water molecules at the air-water interface, firstly we have identified all the water molecules that during the simulation get within 2.5 \AA{} of the WCI. Then, for each water molecule $i$ we construct the characteristic function $h_i(t)$ as a function of its distance from the WCI, $a_i(t)$ :

\begin{equation}
h_i(t) =
\begin{cases}
1 & \quad \text{if } a_i(t) < 2.5 \ \text{\AA{} }\\
0 & \quad \text{if } a_i(t) > 2.5 \ \text{\AA{} }\\
\end{cases}
\label{eq:decorrelation}
\end{equation}

We then calculated the total autocorrelation function by averaging the single molecule correlation function, $C(t) = (TN)^{-1} \sum_i^N \int_0^{T} h_i(\tau + t) h_i(\tau) dt$. 
The same strategy has been used to calculate the path decorrelation function. First, the $h_i^l$ function was constructed for all proton wires $i$ of a certain length $l$, and then the global correlation function for that length $l$ was obtained by averaging, with the appropriate normalization.

\begin{table}
\begin{center}
\caption{Simulations of the excess proton near the air-water surface. In table are reported the 
label, the simulation time and the starting position of the proton with respect to the closest 
surface. In simulation E the distance of the proton from the surface was constrained as in 
\cite{baer2014toward}}
\begin{tabular}{lcc}
\label{tab:tab1}
Label & t [ps] & distance from the closest surface [\AA] \\ \hline
A & 35 & 0 \\
B & 39 & 0 \\
C & 20 & 15 \\
D & 15 & 15\\
E & 20 & 15
\end{tabular}
\end{center}
\end{table}

We are acutely aware that there continues to be an ongoing discussion about the accuracy of DFT 
based simulations to predict the structural, dynamical and spectroscopic properties of bulk 
water\cite{gillanalfemichaelides2016}. Currently, the most accurate potential for neat water that 
reproduces a wide range of these properties across the phase diagram is one built on a many-body 
expansion of the energy\cite{REF4,REF5,REF6}. This model however, is not dissociable and can therefore not account 
for the presence of hydronium and hydroxide ions. DFT based ab-initio  molecular dynamics 
simulations provide a valuable method for understanding many mechanistic details. When using 
DFT, there is always a concern of the sensitivity of results to the quality of the underlying 
electronic structure. Indeed, it has been shown that many of the important mechanistic features for 
proton hopping in processes like recombination of hydronium and hydroxide, are not sensitive to the 
choice the density functional or the basis set employed\cite{HassanaliPrakashEshetParrinello2011}. Moreover,  all of the previous 
theoretical studies described earlier determining PMFs are either based on DFT or empirical 
potentials derived from DFT calculations\cite{baer2014toward,TseLindbergKumarVoth2015}.

\section*{Results}

\subsection*{Proton confinement at the Surface of Water}

Mundy and co-workers recently showed that the PMF to pull the proton from bulk to the air-water interface is flat throughout the slab - in other words, there is no apparent preference for the ion to be at the interface or in the bulk\cite{baer2014toward}. Another more recent study by Voth and co-workers predicted a small propensity, on the order of thermal energy, for the surface\cite{TseLindbergKumarVoth2015}. The 1d-PMFs reported in both these studies have important implications for the underlying dynamics of the proton.  Specifically, if this PMF encodes the slowest degree of freedom, proton motion from the surface to the bulk should involve a tiny barrier, and therefore, one would not expect to see any significant trapping of the proton at the surface. 

\begin{figure}[h!]
\centering
\includegraphics[width=0.425\textwidth]{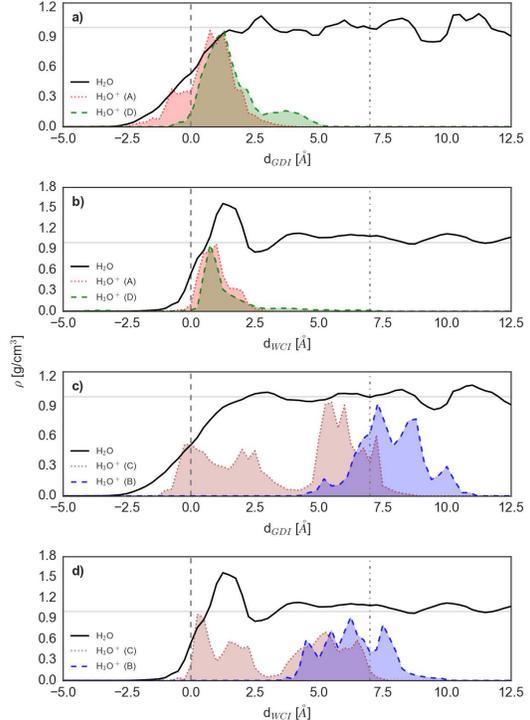}
\caption{ Water density and probability density of finding the excess proton as a function of the distance from the GDI (panel 1 and 3 ) and the WCI (panel 2 and 4 ). In all panels the water density is represented as a solid black line. A gray dashed line indicates the position of the instantaneous or averaged interface, and the starting position of the proton in the sub-surface region in a dot-dashed gray line. The first two panel refer to the proton of simulation A (red, dotted) and simulation B (green dashed) while the lower two panels to simulation C (purple, dotted) and D (ocra, dashed). The protons in simulations A and B do not leave the interface, and they perform rare excursions from their starting point. Proton C leaves its original position and after a few attempts reach the surface while proton D performs excursions in the proximity of its equilibration site, showing a propensity for the bulk. The proton densities are in arbitrary units, and they have been scaled to allow the reader to compare them.}
\label{fig:dens}
\end{figure}

Our simulations paint a very different picture than what the 1d-PMFs suggest. We initiated four independent simulations of the proton, A-D, where A and B were equilibrated with the proton at the surface and C and D with the proton in bulk water (see Methods for more details). In our simulations, no constraints on the structural diffusion of the proton \emph{a-la Grotthuss} are imposed ensuring that the delocalization entropy of the proton is accurately captured\cite{Agmon1995}. In most situations, we find a strong propensity for the proton to be at the surface of water. To see this, we show in Fig. \ref{fig:dens} a) the density distributions of both water and the hydronium ion with respect to the GDI in simulations A and B each of which was run for approximately 35 ps. In both cases, the proton density exhibits a peak close to the GDI indicating much stronger propensity for the surface than what would be expected from earlier 1d-PMF 
calculations.

The distributions shown in Fig. \ref{fig:dens}a average out the instantaneous fluctuations of the interface and consequently the water density quickly plateaus to its bulk value giving no clues into the origins of proton trapping. Fig. \ref{fig:dens} b) shows the density distributions constructed using the WCI interface for A and B (see SI for details on how the WCI interface is constructed). Consistent with previous studies, the water density shows significantly more structuring at the vapor interface.
One might already anticipate that this type of structuring will have an effect on both the thermodynamic and dynamical properties of the proton at the surface of water.

Comparing the distributions obtained with respect to GDI and WCI, we see that in the former, we cannot distinguish between the first and second solvation layers. Using WCI, we see that the peak in the proton density for both A and B lies very close to the first water peak in the first solvation layer.  In fact, the proton density is peaked slightly more toward the vapor phase than the first peak of water. Besides locating the peak positions, the fluctuations of the proton distributions are also quite different. In both A and B, we see that the proton density is much broader using GDI - in A, this density covers a spread of about 5 \AA{} at the interface, while with WCI, this is narrower by a factor of about 2. This effect is also observed in simulation B where the proton is completely confined at the surface. In this case, the GDI proton density features a shoulder in the bulk region while with WCI this feature becomes a lot less pronounced. Although we cannot make any quantitative conclusions from these results, it is clear that the instantaneous density fluctuations have an effect on its underlying PMF, and alter the barrier and pathway for its transfer from the surface into the bulk.

The protons in simulations C and D that were initiated in the bulk, provide more insights into the effective PMF that the proton experiences as it approaches the surface. The proton density in simulation C is characterized by two peaks, one in the bulk between 5-7.5 \AA{} and another at the surface like in A and B, between 0-2.5 \AA{}. The origin of this is because the proton, initially in the bulk region, moves towards the surface during the simulation. Simulation D, on the other hand, features density peaks located exclusively in the bulk region.  It is clear once again that there are appreciable differences between the proton distributions obtained with respect to GDI and WCI. In both C and D, the entire proton distributions in the bulk region shift closer to the interface and in the former, lead to a more pronounced presence of the proton at the surface. The origin of this behavior can be rationalized because the proton which is far from the GDI, is in fact much closer to the WCI due to the significant fluctuations of the latter.

Simulations A through D reveal that the proton has a depleted density in the region between 2.5-5.0 \AA{}. The difference in the water density as inferred from the WCI metric provides clues into the origin of this effect. The first minimum of this distribution occurs roughly at $\sim$ 2.5 \AA{} which coincides with the region of lower water density in all the simulations. The evidence built here thus suggest that the instantaneous density fluctuations play a critical role in affecting the potential that the proton experiences at the air-water interface. 

Although the proton is confined at the interface, it is not immobile and undergoes Grotthuss-like diffusion. In bulk, it was recently shown that this process involves the motion of the proton through an architecture of proton-wires in the hydrogen bond network\cite{HassanaliGibertiCunyKuhneParrinello2013}. To probe the effects of the structuring we will next examine the properties of water wires at the interface.

\subsection*{ Water Wires at the Air-Water Interface}

Bulk liquid water at ambient conditions is a percolating network made up of local directional correlations between water molecules which create the underlying architecture for water wires\cite{HassanaliGibertiCunyKuhneParrinello2013}. These are essentially hydrogen paths connecting different water molecules along
which the proton can diffuse. Since the air-water interface features a peculiar structuring of water molecules, we wanted to understand how the water wires at the surface could facilitate proton confinement at the surface. If one is to imagine Grotthuss-type proton shuttling involving either stepwise or concerted proton hopping over several hydrogen bonds, this analysis aims at quantifying how deep away from the WCI successful transfer events along the proton wire would take the proton.   

\begin{figure}[h!]
\centering
\includegraphics[width=0.425\textwidth]{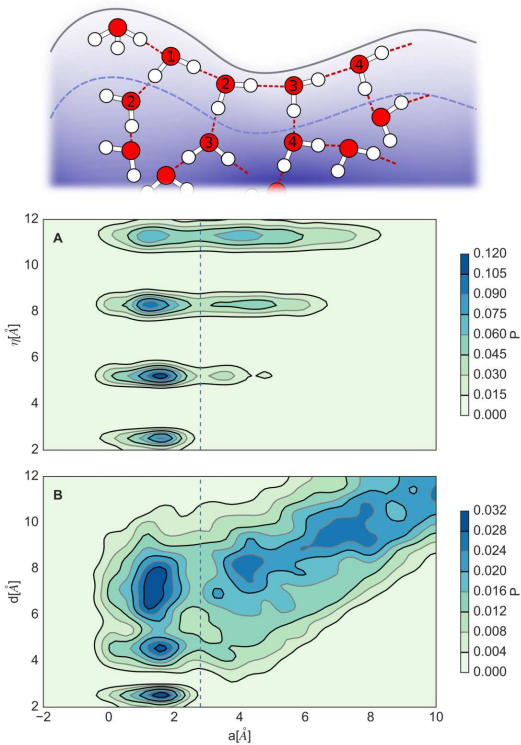}
\caption{Panel A) shows the probability densities of finding a directed pathways between the hydronium ion and a water molecule at distance $d$ and located at d$_{WCI}$ from the WCI. Most of the wire run parallel to the surface, and will favor the lateral diffusion of the excess proton. To penetrate in the slab, the hydronium ion has to cross a low probability region corresponding to the first minimum in the density profile from the WCI (dashed blue line). Panel B) shows the likelihood of a wire composed of one, two, three and four water molecules to be parallel or perpendicular to the surface. The probability of finding a wire that runs parallel to the surface is higher than perpendicular, for wires composed of one, two and three water molecules. The same minimum in the distribution observed in panel A) can also be noticed in panel B), in correspondence of the minimum in the water density from the WCI.}
\label{fig:pathp1} 
\end{figure}

We begin by illustrating the network surrounding the proton confined at the interface in simulation A.
To probe the network, directed shortest paths emanating from the water molecule hosting the proton were determined. To this extent, we introduce two different probability distributions, P($\eta$,d$_{WCI}$) and P($d$,d$_{WCI}$). The variable $\eta$ is the sum of all oxygen-oxygen distances forming the shortest geodesic path from the proton and the last molecule along the path, $d$ is the distance between the proton and the last water molecule along the path and finally $d_{WCI}$ is the distance of this last water molecule from the WCI interface. The distributions P($\eta$,d$_{WCI}$) and P($d$,d$_{WCI}$) are shown in panels a and b of Fig. \ref{fig:pathp1}. The blue dashed line highlights the position of the first minimum of the WCI density profile. We see that for paths up to length three (where $\eta$ is less than 9 \AA{}), there is a strong preference for the wire to remain within the first layer of the interface. Proton jumps over a single hydrogen bond will always keep it confined at the interface. As one move to paths of length two, three and four, the distribution of pathways broadens resulting in paths that can dig deeper away from the surface although there is still a strong preference for longer paths to remain within the first water layer.  

The proton wires do not form linear chains - in fact, they are naturally quite coiled. To appreciate this a bit better, Fig. \ref{fig:pathp1} b) illustrates the distribution P($d$,d$_{WCI}$), which shows that paths of length greater than two tend to exhibit more coiled character and merge into a broader basin for $d$ greater than 6 \AA{}. As a reference, for perfectly linear proton wires there would be distinct populations in the P($d$,d$_{WCI}$) distribution at values of $d$ at approximately 3 \AA{} and then subsequent increments of the same length. See the Supporting Information for more details on quantifying the coiled nature of the wires. Despite being coiled, these proton wires still have a strong preference to orient parallel to the surface. The feature of parallel proton wires is not unique to the environment surrounding the excess proton. Interestingly, the features elucidated in Fig. \ref{fig:pathp1} are also characteristic of neutral water molecules at the surface (see SI for similar distributions constructed for neutral water molecules). These observations are consistent with Willard and Chandlers analysis of the surface of water where they found a tendency for water molecules to form hydrogen bonds within the same solvation layer\cite{WillardChandler2010}. The parallel water wires are a manifestation of a longer-range ordering effect caused by these local hydrogen bonding patterns.

\begin{figure}[h!]
\centering
\includegraphics[width=0.425\textwidth]{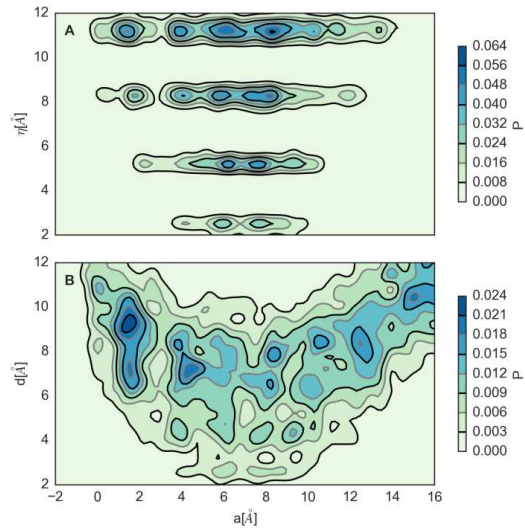}
\caption{Panel A) shows the probability densities of finding a directed pathways between the hydronium ion and a water molecule at distance $d$ and located at d$_{WCI}$ from the WCI. If the proton is buried in the slab, the probability of performing excursion by increasing, decreasing or keeping the same distance from the surface is the roughly the same.  Panel B) shows the likelihood of a wire composed of one, two, three and four water molecules to be parallel or perpendicular to the surface. The probability of finding a wire that runs parallel to the surface is higher than perpendicular, for wires composed of one, two and three water molecules. The same 
minimum in the distribution observed in panel A) can also be noticed in panel B), in correspondence of the minimum in the water density from the WCI.}
\label{fig:pathp4D}
\end{figure}

In order to appreciate better how the topology of the hydrogen bond network forming the proton wires changes for situations where the proton is not confined at the interface, we also examined the probability distributions shown in \ref{fig:pathp1}A) and B) for simulation D. The resulting probability distribution is illustrated in Fig.\ref{fig:pathp4D}. Here we observe that the proton wire distributions are naturally very different in that there are now two \emph{wings} - proton wires going to the surface and those going into the bulk. One observes that for the proton undergoing diffusion parallel to the surface normal, there are proton wires of length 3 and 4 with distinct populations close to the surface and also into the bulk region. This is somewhat expected since there are now proton wires that begin at the proton and can end up either at the surface as well as percolate into the bulk. It should be mentioned that the structure in the distributions shown in Fig.\ref{fig:pathp4D} going into the bulk, in part reflect the short timescales of the ab initio simulations. We expect that on much longer timescales this structure to be averaged out. The preceding analysis thus clearly shows that an understanding of the dimensionality of the random walk that the proton experiences is a critical part of understanding its propensity for the air-water interface.

\subsection*{Water Defects at the Air-Water Interface} 

\begin{figure}[ht]
\centering
\includegraphics[width=0.45\textwidth]{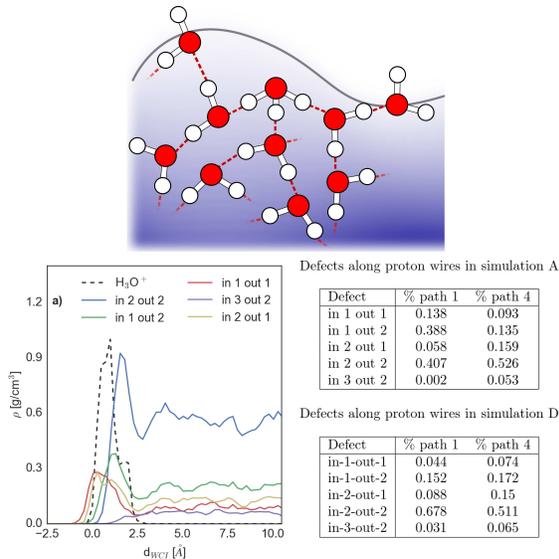}
\caption{Defects concentration as a function of the distance from the instantaneous water interface. The probability density of the proton is depicted in a dashed black line. The probability densities of the principal defects observed at the instantaneous interface are reported in different colors. Each topological motifs have an increase in concentration at the interface but the in 3 out 2 over-coordinated water molecules. The peak of each topological motif at the interface is located at different depth, meaning that the first layer from the WCI is also structured, a description of the structure and a rationalization of the position of the proton is explained in the text.}
\label{fig:defects} 
\end{figure} 

The existence of the proton wires is not the only factor that can contribute to arresting the proton at the interface. In bulk water, for example, specific changes in the local coordination patterns of the proton and surrounding water molecules play an important role in either facilitating diffusion or trapping it in the network\cite{Agmon1995,LapidAgmonPetersenVoth2005,BerkelbachLeeTuckerman2009,HassanaliGibertiCunyKuhneParrinello2013}. Since the oxygen atom of the water hosting the proton is a weak acceptor of a hydrogen bond and thus hydrophobic, one might anticipate that undercoordinated water molecules at the interface are good candidates for accommodating the proton. Given the highly structured water at the interface as seen in Fig. \ref{fig:dens}, it is interesting to examine how this region is built up of water molecules with different coordination patterns or water defects\cite{KuoMundy2004,henchman2010,GasparottoHassanaliCeriotti2016} and more specifically, their implications on proton trapping at the interface.

Fig. \ref{fig:defects} shows the concentration of several important defects with respect to the WCI for simulation A: right at the interface close to the vapor phase and above the average proton position, there is a high concentration of 1in-1out and 2in-1out defect water molecules which essentially correspond to water molecules with dangling O-H bonds pointing to the vacuum. Just below the peak position of the proton, there is a higher concentration of 1in-2out defects and deeper in, but still, within the first structured layer, lie the tetrahedral 2in-2out water molecules. This feature is observed in all other simulations (see the Supporting Information for simulations B through D) and reflects the underlying structuring of the water at the interface and its subsequent effects on where the proton is most stabilized. 

To understand better how these coordination defects populate the environment of the proton, we examined the identity of the coordination defects of the water molecules along the proton wires. The top table in Fig.\ref{fig:defects} shows that the 1in-2out and 2in-2out water defects 
have the highest probability of being the nearest neighbor of the proton in simulation A where the proton is confined at the surface.  As one moves three water molecules away along a proton wire, the probability of finding a 2in-2out increases with a corresponding decrease in the 1in-2out population. In simulation D where the proton is diffusing in the bulk, we observe that the proportion of 1in-2out water molecules as the first neighbor is reduced compared to simulation A. In both
simulations A and D, there is a small probability of finding a 1in-1out or 2in-1out water defect along the wire.

These features elucidated in the preceding discussion, reflect the amphiphilic nature of the proton - on the one hand, the hydrophobic oxygen positions itself as close as possible to the WCI interface (within 1 \AA{}) to maximize its contact with the vapor phase, as already observed in other calculations\cite{WuChenWangPaesaniVoth2008,HassanaliGibertiSossoParrinello2014}. On the other hand, the character of the first neighbor of the proton must be conducive to accepting a strong hydrogen bond. The 1in-1out and 2in-1out water defects which are pinned right at the WCI are not favorable candidates. This is because they have an O-H bond that is dangling and pointing the vacuum and are hence not permissive to allowing for collective proton excursions along the hydrogen bonds of the proton wires. The 1in-2out water molecules are perhaps the best candidates for either proton conduction or hosting the proton, and their prominence for wires of length one implies that they mostly accept a hydrogen bond exclusively from the proton.

\subsection*{Dynamics of HB Network at the Air-Water Interface}

So far, we have illustrated that the presence of the air-water interface results in the formation of coordination defects that can facilitate the confinement of the proton at the surface as well as directed proton-wires along which the protons migrate.  To understand the origins of the proton trapping, it is important to assess the dynamical processes involved in the reconstruction of the hydrogen bond network.

\begin{figure*}[ht!]
\centering
\includegraphics[width=\textwidth]{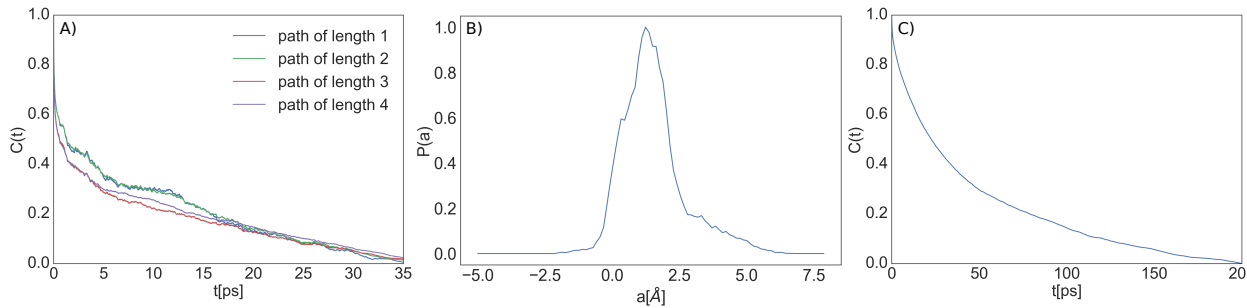}
\caption{Panel A) Correlation function of directed pathways of length 1 to 4. It is remarkable to 
notice that the directed pathways at the surface can have a very long lifetime, which may span from 
5 to 20 ps. Most of the water molecules are trapped in the first layer on the length-scale 
of our simulations, and do not diffuse in the bulk, as can be seen from the histogram of their 
position in panel B). As a reference, we also reported the correlation function for a classical 
system composed of 384 SPC/E water molecules in panel C). }
\label{fig:dynamics} 
\end{figure*}

In bulk water, numerous studies have shown that proton motion is coupled to pre-solvation - a concerted and collective fluctuation of water molecules in the environment of the proton which facilitates its diffusion\cite{markovitch2008special,BerkelbachLeeTuckerman2009,HassanaliGibertiSossoParrinello2014}. In the context of this work, it is interesting to examine the timescales associated with the fluctuations of water in and out of the first layer of the WCI as this could potentially lead to the changes in the proton wires that would take the proton from the surface into the bulk. On the timescales that are accessible with AIMD, a large proportion of the water molecules remain trapped in the first layer of the WCI. To provide the reader a qualitative idea of the slow density relaxation of the surface, we examined the probability distribution of the distance traveled by water molecules that begin within the first layer of the air-water interface relative to the WCI. This is shown in Fig \ref{fig:dynamics} A). We see here that on the short timescale of 35 ps of simulation affordable by AIMD, most of the water molecules remain within at the air-water interface.

In order to quantify better the slow exchange time, we also computed a residence time correlation function from a longer simulation of the air-water interface using an empirical potential TIP4P/EW. This correlation function essentially measures the average time a water molecule resides within 2.5 \AA{} of the WCI (see Methods for more details). This relaxation dynamics is shown by the solid red curve in Fig.\ref{fig:dynamics}B). The residence time correlation function can be fit to a double exponential involving two timescales: 9.009 and 62.5 ps - the faster timescale is consistent with that reported in an earlier study\cite{kuhne2011}. Our simulations show that after about 40ps, 40\% of the water molecules still remain within the first layer and confirm that there is indeed slow residence time exchanges of water molecules in and out of the air-water interface.

Despite the long residence times of water molecules, the network undergoes reconstruction. To quantify this, we also examined the relaxation dynamics of proton wires of different lengths shown in Fig. \ref{fig:dynamics}C). Qualitatively, we see that $\approx$ 20 \% of the pathways have a lifetime of 5 ps, and $\approx$ 10 \% can survive up to 15-20 ps. The relaxation dynamics was fit to two exponentials involving relaxation timescales of $\approx$ 1.7 and 40 ps on average (see the SI for fits). The factors contributing to the relaxation of the directed wires include both proton diffusion along the water wires and the breakage or formation of hydrogen bonds within the first layer. The slow relaxation of the wires is consistent with theoretical studies calculating 2D-SFG spectra that conclude where hydrogen bond switching was found to be three times slower than in the bulk\cite{YiCunGruenbaumSkinner2013}. The slow relaxation timescales we observe at the interface is also in line with previous simulations exploring interface dynamics at both the air-water and metal-water interfaces\cite{PaulChandra2004}.

All in all, we find that the relaxation dynamics of hydrogen bond network at the interface is characterized by an interesting feature of slow residence exchange times between the water molecules within the first layer of the WCI and the bulk. Similar features have also been observed for water around biological systems like proteins\cite{kumarpeonbagchizewail2002,Lihassanalikaozhongsinger2007,lihassanalisinger2008}. Network reconstruction still occurs at the surface allowing for proton diffusion to occur along the proton wires that run predominantly parallel to the surface. 

\subsection*{Grotthuss Mechanism at the Air-Water Interface}

The preceding analysis has shown that the instantaneous density fluctuations as probed through the WCI, has non-trivial implications on the propensity of the proton to be trapped at the surface of water. One of the essential ingredients leading to this is that longer range structuring leads to very specific features in the water wires that exist in the first water layer as well as the concentration of topological coordination defects in the vicinity of the proton. In the ensuing analysis, we elucidate in a more anecdotal way, the underlying mechanism by which protons move through these wires and how this process couples intimately with the proximity of the proton to the WCI.

\subsubsection*{Simulations A and B} 

\begin{figure}[ht]
\centering
\includegraphics[width=0.5\textwidth]{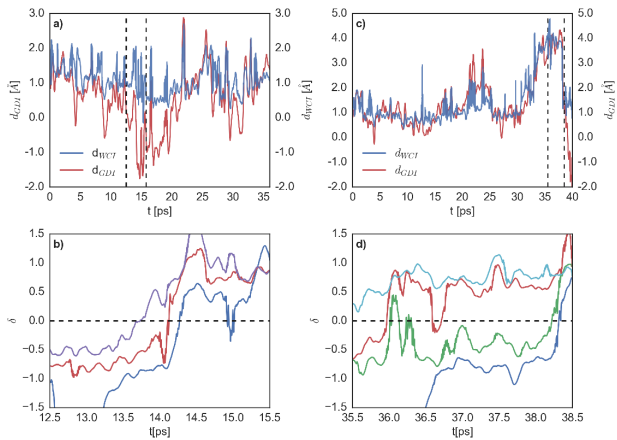}
\caption{Time evolution of the excess proton in simulations A and B. Panel a )illustrates the position of the excess proton from the GDI (red) and the WCI (blue). During the simulation, the proton migrates from a "valley" to the top of a "wave"  with a triple concerted proton transfer, illustrated using the proton transfer coordinate $\delta$ in panel b). A snapshot of the water wire along which the transfer occurs is shown in fig \ref{fig:wireA}.  In panel c) we reported the evolution of the proton in simulation C. After a brief transient inside the slab, the proton migrates at the surface with a series of concerted and stepwise transfers illustrate in panel d). } 
\label{fig:conc-surf1} 
\end{figure}

\begin{figure}[ht]
\centering
\includegraphics[width=0.5\textwidth]{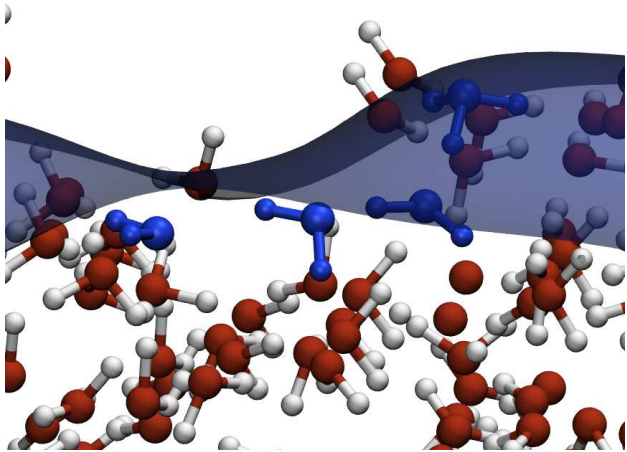}
\caption{Snapshot of the water wire along which the triple concerted proton transfer occurs in simulation A. The water molecules composing the wire are highlighted in blue, with the hydronium residing on the highest one. The blue transparent surface is a pictographic representation of the interface.}
\label{fig:wireA}
\end{figure}

The proton trapped in the first layer of water in simulations A and B exhibits very rich dynamical behavior. Although we cannot converge any dynamical properties like a diffusion constant, we see that the mean square displacement (MSD) is quite different for the two situations - the total MSD for A is about ten times that of B (see SI for details). This type of variation in the MSD has also been observed for proton diffusion in the bulk\cite{HassanaliGibertiCunyKuhneParrinello2013} and reveals the challenges of converging dynamical properties from these types of simulations. To gain insights into the molecular origins of these effects, we examined the position of the proton with respect to the GDI (d$_{GDI}$) and the WCI (d$_{WCI}$) for simulation A, reported in panel a) of Fig \ref{fig:conc-surf1}. Qualitatively, the motion of the proton with respect to both parameters appear to be dynamically correlated insofar as over the course of the first 10 ps, the distance from the interface of the proton decreases. This motion has larger fluctuations in the GDI metric compared to the WCI, consistent with the proton density distributions shown earlier. Between 12 and 15 ps, an interesting event occurs whereby d$_{GDI}$ decreases by about 3 \AA{}.  The origin of this comes from a concerted motion of three protons along a water wire running parallel to the water surface as seen in the PTC coordinates reported in panel 2.  During this activity, however, the distance from the WCI does not change much. Observing the proton position exclusively with respect to the GDI would lead to the erroneous interpretation that the proton is \emph{spilling} above the surface. However, the distance from the WCI illustrates that the proton essentially \emph{surfs} along the instantaneous interface. Fig \ref{fig:wireA} shows a snapshot of the simulation illustrating the instantaneous interfaces as well as the proton wire.

The proton in simulation B although trapped at the interface like A, exhibits rather different dynamical behavior. As indicated earlier, its MSD is significantly lower than that in A (see SI).  For the first 15 ps, the proton remains localized on one water molecule without much activity. Between 20-25 ps the proton rattles between different water molecules, modulating its position along d$_{WCI}$ and d$_{GDI}$. This rattling does not lead to any successful proton transfer events. Between 30 and 35 ps, d$_{WCI}$ and d$_{GDI}$ increase by about 3-4 \AA{} and the proton moves into the sub-surface region. This process occurs through a combination of closely spaced stepwise and concerted proton hops. After about 2 ps of residing in this region, a double-concerted proton hopping event (at around 38.5 ps)  makes the proton \emph{float} back right to the top of the water surface from where it began.

\subsubsection*{Simulations C and D}

Simulations C and D were equilibrated and initialized in the bulk region of the slab. As seen in the density plots earlier, C exhibits peaks in the density both in the bulk and at the interface, whereas D is localized in the bulk. We begin with simulation C: for the first eight ps, the proton resides in the subsurface region between 5.0 and 7.5 \AA{} from the GDI (see panel a) of Fig \ref{fig:conc-surf1}). In these early stages, d$_{GDI}$ shows a more pronounced drift toward the interface compared to $d_{WCI}$. Between 8-12 ps there is a large change in both d$_{WCI}$ and d$_{GDI}$ - there is a sharp rise and then decrease by several angstroms during which the proton makes its way to the surface. This migration takes place from 8 to 16 ps and is achieved through a mixture of concerted and stepwise proton transfer events as seen in panel b).

\begin{figure}[ht]
\centering
\includegraphics[width=0.5\textwidth]{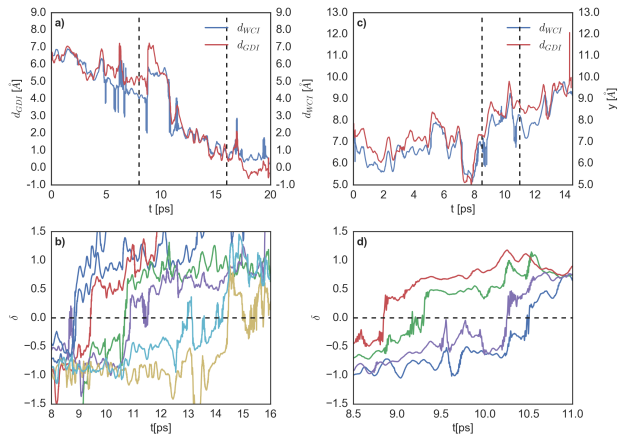}
\caption{Time evolution of the excess proton in simulations C and D. Panel a) illustrates the time evolution of the excess proton with respect to the GDI (red) and the WCI (blue). After 8 ps, the proton ascends with a series of concerted and stepwise transfers up to the air-water interface. The proton transfer coordinates of the ascending process are reported in panel b). The evolution of the proton in simulation D is reported in panel c). For the entire simulation, the proton resides inside the slab and drift slightly to its center of mass. Despite being trapped along the y coordinate, however, it performs concerted proton transfer exploration in the xz plane. In panel d) has been reported an example of such transfer using the proton transfer coordinates along which the shuttling occurs.}
\label{fig:conc-surf2} 
\end{figure}

In sharp contrast to C, simulation D is the only case where the proton remains in the sub-surface region between 4 to 10 \AA{} from the WCI for the entire simulation (see Fig \ref{fig:conc-surf2} c)). The MSD behavior for simulations C and D is quite similar over the course of the 20 ps (see SI), meaning that the proton is undergoing lateral diffusion the xz plane as was observed in previous studies\cite{WuChenWangPaesaniVoth2008}. As for the other simulations, the majority of the transfer occurs in a concerted manner, while the minority of them involves the shuttling of only one proton. In Fig. \ref{fig:conc-surf2} d) we illustrate an example where over the course of 2 ps, there are four successful proton transfer events. 

\subsection*{Discussion and Conclusions} 

Despite its apparent simplicity, the very basic fundamental question of how attracted protons and hydroxide ions are to the air-water interface
remains a highly contentious topic from both experimental and theoretical
fronts. In this work, we revisited this problem by focusing on the affinity of the proton for the surface of water using DFT-based AIMD simulations. We do not claim to resolve this controversy by any means, but instead, shed
new insights into aspects of the problem that have been previously neglected.
The one-dimensional PMFs built on the position of the proton relative to the GDI, do not capture the complexity in both the thermodynamic and dynamical aspects of the proton affinity for the surface or its Grotthuss diffusion at the interface. One of the most important and central findings of this work is that the instantaneous fluctuations of the interface are an essential ingredient for understanding both the physics and chemistry of this
system.

The emerging evidence built on this work, suggests that the structuring of the first layer as revealed by the WCI analysis has important implications on the underlying potential that the proton feels at the interface. The Gibbs dividing surface is a construction of an interface that averages out
instantaneous fluctuations and thus also its realistic corrugations.
Subsequently, it features in a specious way, more pronounced fluctuations of the proton position from the surface. 
For the situations in our simulations where the proton is confined at the surface, their apparent fluctuations from the surface are reduced. In essence, the large fluctuations of the WCI which penetrate deeper into the bulk region, enhance the \emph{presence} of the proton at the interface.

The highly structured water at the interface is rooted in various properties of the hydrogen bond network of water.
The preference for water molecules to hydrogen bond to
each other within the first layer\cite{WillardChandler2010},
leads to the formation of proton wires running parallel to the surface making the release of the proton into the bulk more difficult. This is curiously similar to what has been observed for water near hydrophobic membrane surface where the delay of surface-to-bulk transfer facilitates a pathway for lateral proton diffusion\cite{Pohl2012}.

The dynamical processes at the air-water interface involving the reconstruction of the hydrogen bond network feature a range of different timescales. One of the important characteristics is the rather long residence time of water molecules at the interface. We believe that this feature is important for the creation of long-lived proton wires running parallel to the interface which makes the release of the proton into the bulk difficult. It would be interesting in the future to examine in detail the effect of the structured water at the interface on the hydroxide ion.

\section*{Acknowledgments}
The Author would like to thank Prof. Michele Ceriotti for insightful discussion on the subject. The work performed by FG was funded by was supported by MICCoM, as part of the Computational Materials Sciences Program funded by the U.S. Department of Energy, Office of Science, Basic Energy Sciences, Materials Sciences and Engineering Division, under grant DOE/BES 5J-30161-0010A.

\bibliography{main}

\bibliographystyle{unsrt}

\end{document}